\documentclass[aps,prl,reprint,groupedaddress,amssymb,citeautoscript]{revtex4-2}

\usepackage{amsmath}
\usepackage{xcolor}
\usepackage{natbib}
\usepackage{graphicx}
\usepackage{hyperref}
\usepackage[normalem]{ulem} 
\hypersetup{
    colorlinks=true,
    linkcolor=blue,
    filecolor=magenta,
    urlcolor=cyan,
}

\begin{document}


\title{Coexistence of local magnetism and superconductivity in the heavy-fermion CeRh$_2$As$_2$ revealed by $\mu$SR studies
}


\author{Seunghyun Khim}
\email[]{Seunghyun.Khim@cpfs.mpg.de}
\affiliation{Max Planck Institute for Chemical Physics of Solids, No\"thnitzer Stra{\ss}e 40, 01187 Dresden, Germany}

\author{Oliver Stockert}
\email[]{Oliver.Stockert@cpfs.mpg.de}
\affiliation{Max Planck Institute for Chemical Physics of Solids, No\"thnitzer Stra{\ss}e 40, 01187 Dresden, Germany}

\author{Manuel Brando}
\affiliation{Max Planck Institute for Chemical Physics of Solids, No\"thnitzer Stra{\ss}e 40, 01187 Dresden, Germany}

\author{Christoph Geibel}
\affiliation{Max Planck Institute for Chemical Physics of Solids, No\"thnitzer Stra{\ss}e 40, 01187 Dresden, Germany}

\author{Chirstopher Baines}
\affiliation{Laboratory for Muon Spin Spectroscopy, Paul Scherrer Institute, CH-5232
Villigen PSI, Switzerland}

\author{Thomas J. Hicken}
\affiliation{Laboratory for Muon Spin Spectroscopy, Paul Scherrer Institute, CH-5232
Villigen PSI, Switzerland}

\author{Hubertus Luetkens}
\affiliation{Laboratory for Muon Spin Spectroscopy, Paul Scherrer Institute, CH-5232
Villigen PSI, Switzerland}

\author{Debarchan~Das}
\affiliation{Laboratory for Muon Spin Spectroscopy, Paul Scherrer Institute, CH-5232
Villigen PSI, Switzerland}

\author{Toni Shiroka}
\affiliation{Laboratory for Muon Spin Spectroscopy, Paul Scherrer Institute, CH-5232
Villigen PSI, Switzerland}

\author{Zurab Guguchia}
\affiliation{Laboratory for Muon Spin Spectroscopy, Paul Scherrer Institute, CH-5232
Villigen PSI, Switzerland}

\author{Robert Scheuermann}
\affiliation{Laboratory for Muon Spin Spectroscopy, Paul Scherrer Institute, CH-5232
Villigen PSI, Switzerland}



\date{\today}

\begin{abstract}

The superconducting (SC) state ($T_\mathrm{c}$ = 0.3~K) of the heavy-fermion compound CeRh$_2$As$_2$, which undergoes an unusual field-induced transition to another high-field SC state, emerges from an unknown ordered state below $T_\mathrm{o}$ = 0.55 K.
While an electronic multipolar order of itinerant Ce-4$f$ states was proposed to account for the $T_\mathrm{o}$ phase, the exact order parameter has not been known to date.
Here, we report on muon spin relaxation ($\mu$SR) studies of the magnetic and SC properties in CeRh$_2$As$_2$ single crystals at low temperatures.
We reveal a magnetic origin of the $T_\mathrm{o}$ order by identifying a spontaneous internal field below $T_\mathrm{o}$ = 0.55~K.
Furthermore, we find evidence of a microscopic coexistence of local magnetism with bulk superconductivity.
Our findings open the possibility that the $T_\mathrm{o}$ phase involves both dipole and higher order Ce-4$f$ moment degrees of freedom and accounts for the unusual non-Fermi liquid behavior.

\end{abstract}


\maketitle


\textit{Introduction} - CeRh$_2$As$_2$ (with a critical temperature $T_\mathrm{c}$ of 0.3~K) has been attracting attention due to its unique two-phase superconductivity.
This distinct behavior is highlighted by a field-induced transition from a low-field even-parity superconducting (SC) state to a high-field odd-parity state \cite{khim2021,Landaeta2022}.
The crystal structure of CeRh$_2$As$_2$ lacks local inversion symmetry at the Ce site while being globally centrosymmetric \cite{Madar1987}.
This nonsymmorphic structure, characterized by an inversion pair of the Ce sites in the unit cell, leads to sublattice degrees of freedom \cite{Fischer2023}.
Hence, superconductivity in this compound is expected to be influenced by this structural feature and associated antisymmetric spin-orbit coupling \cite{Anderson1984,Yoshida2012,Maruyama2012,Sigrist2014}.
Indeed, the SC phase diagrams of CeRh$_2$As$_2$ have been well explained by a schematic theoretical model which describes a stack of weakly-linked layered superconductors with strong spin-orbit interactions \cite{khim2021,Schertenleib2021,Mockli2021,Landaeta2022}.
However, a comprehensive microscopic picture that includes also the role of the local Ce-4$f$ states and Kondo physics still remains elusive \cite{Nogaki2022}.
A common observation of multiphase superconductivity in heavy-fermion compounds that lack local inversion symmetry, such as UPt$_3$ \cite{Huxley2000}, UBe$_{13}$ \cite{Heffner1990}, UTe$_2$ \cite{Sakai2023}, and CeSb$_2$ \cite{Squire2023}, may suggest a potential for a more general explanation among $f$-electron systems \cite{Nica2022,Hazra2023}.

\begin{figure*}
\centering
\includegraphics[width=160mm]{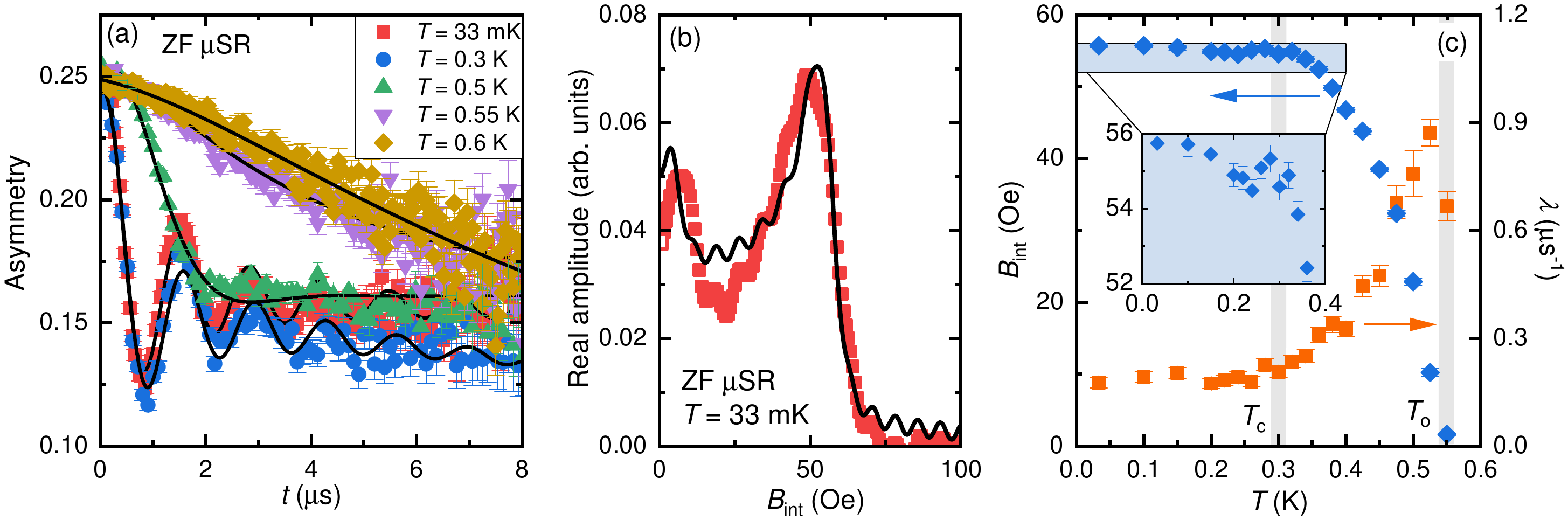}
\caption{(a) Zero-field $\mu$SR time spectra for CeRh$_2$As$_2$ obtained at 0.033, 0.3, 0,5, 0.55, and 0.6 K, respectively. (b) Real part FFT of the time spectra at $T$ = 33 mK. The solid black lines in (a) and (b) are the fits to the data modeled by Eq. \ref{eq:Bessel}. (c) Temperature dependences of the relaxation rates $\lambda$ (orange squares) and the internal field $B_\mathrm{int}$ = 2$\pi \nu$/$\gamma_{\mu}$ (blue diamonds) obtained from the fitting. The inset emphaizes the behavior of $B_\mathrm{int}(T)$ near $T_\mathrm{c}$.}
\label{fig:FIG1}
\end{figure*}

More intriguingly, the SC state of CeRh$_2$As$_2$ emerges from an unknown ordered phase below $T_\mathrm{o}$ = 0.55~K.
A possible electronic origin of the $T_\mathrm{o}$ order was suggested based on (1) the absence of an anomaly by bulk magnetic probes \cite{khim2021}, (2) the quasi-quartet crystal electric field (CEF) configuration for the local Ce state \cite{Jeevan2006,Christovam2024}, and (3) the highly anisotropic and multiphase magnetic phase diagrams of $T_\mathrm{o}$($B$) which is reminiscent of a quadrupolar order in the Ce-based quartet-CEF systems \cite{Effantin1985}.
This led to the proposal of a quadrupolar density wave (QDW) state, realized by an elaborate combination of the quasi-quartet CEF and Kondo interactions \cite{Hafner2022,Semeniuk2023}.
On the other hand, nuclear quadrupole resonance (NQR) studies show hints of an antiferromagnetic (AFM) order with $T_\mathrm{N}$ $\sim$ 0.3~K very close to $T_\mathrm{c}$ \cite{Kibune2021}, and nuclear magnetic resonance (NMR) measurements claimed that this AFM state coincides with the low-field SC state \cite{Ogata2023}.
Heat-capacity measurements also detected an anomaly just below $T_\mathrm{c}$, implying a possible AFM order \cite{Chajewski2024}.
Further experimental characterization is therefore required to identify the precise nature of the highly cooperative phenomena taking place at $T_\mathrm{o}$ and $T_\mathrm{N}$.
This will be an essential prerequisite for understanding the microscopic SC pairing mechanism.
Moreover, we cannot completely exclude the alternative scenario for two-phase superconductivity, in which a field-induced change in additional orders  intertwined with superconductivity would drive the field-induced SC transition \cite{Machida2022}.

In this work, we report on zero-field and transverse-field muon spin relaxation ($\mu$SR) studies of the magnetic and SC properties in quality-improved CeRh$_2$As$_2$ single crystals at low temperatures.
We observe a spontaneous internal field below $T_\mathrm{o}$ = 0.55~K, revealing a magnetic origin of the $T_\mathrm{o}$ order.
Furthermore, we find evidence of a microscopic coexistence of the magnetic order and bulk superconductivity below $T_\mathrm{c}$ = 0.3~K.
In combination with experimental results from different magnetic probes, we discuss a possible dynamic nature of the magnetic order.
Our findings show an unusual nature of the $T_\mathrm{o}$ state involving both dipole and higher-order Ce-4$f$ moment degrees of freedom.

\textit{Experiments} - Single crystals of CeRh$_2$As$_2$ were grown by the Bi-flux method \cite{khim2021}.
Zero-field (ZF) and weak transverse-field (wTF) $\mu$SR experiments were performed on the FLAME instrument at Paul Scherrer Institute (PSI, Villigen, Switzerland).
Heat capacity measurements were carried out to determine the transition temperatures of $T_\mathrm{c}$ and $T_\mathrm{o}$ (see Supplemental Materials \cite{SM}).
We note that these crystals have an improved quality compared to those used in the previous studies \cite{khim2021,Semeniuk2023}.
A total of nine single crystals of millimeter size were co-aligned and mounted on a copper plate.
The sample temperature was regulated in a dilution refrigerator from $T$ $\sim$ 30~mK to 2~K.
The crystals were mounted with the tetragonal $c$ axis parallel to the incident $\mu^{+}$ beam direction.
The muon beam was longitudinally polarized for the ZF measurements, in which the time-dependent muon spin polarzation is monitored by the asymmetric position counts between the forward and backward detectors.
For the wTF measurements, magnetic fields were applied parallel to the $c$ axis with the muon transversely polarized and the asymmetry in the counts between the left and right detectors is recorded in the time domain.
The $musrfit$ software provided by PSI \cite{musrfit} was used to fit the muon time spectra.

\textit{Results and discussions} - Figure \ref{fig:FIG1}(a) shows ZF-$\mu$SR time spectra at low temperatures.
The muon asymmetry shows a weak depolarization in the paramagnetic normal state at $T$ = 0.6~K.
As the temperature is lowered, the slow relaxation turns into a rapid damping at $T$ = 0.5 K and a clear oscillation appears at lower temperatures.
This finding clearly demonstrates a spontaneous and coherent internal field established below $T_\mathrm{o}$.
As shown in Fig. \ref{fig:FIG1}(b), the internal field distribution, estimated from the real part of the Fast Fourier Transformation (FFT) of the time spectra at $T$ = 33 mK, exhibits a single pronounced peak, which corresponds to an internal field of $\sim$ 50 Oe.
This implies that the signal arises mainly from a single muon site.
We note that this is predominantly an in-plane component of the internal field probed by the $c$-axis polarized muons.

The evolution of the muon spectra was analyzed by a phenomenological model that includes the Bessel function with exponential decay, written as
\begin{equation}
A(t) = A_{0} \left[ \alpha J_0 \left( 2\pi\nu t + \frac{\pi\phi}{180} \right) e^{-\lambda t} + (1-\alpha)e^{-\lambda_{0} t} \right],
\label{eq:Bessel}
\end{equation} 
where $\alpha$ is the fraction of the oscillating contribution, $\nu$ and $\phi$ are the oscillation frequency and the phase shift, and $\lambda$ and $\lambda_0$ are the relaxation rates for the oscillating and non-oscillating parts, respectively.
Although the choice of the Bessel function is empirical, this model reasonably reproduces the experimental data, as seen in Fig. \ref{fig:FIG1}(a).
The temperature dependences of the obtained relaxation rates and internal field $B_\mathrm{int}$ = 2$\pi \nu$/$\gamma_{\mu}$, where the gyromagnetic ratio of the muon $\gamma_\mu$/2$\pi$ is 135.5 MHz/T, are summarized in Figure \ref{fig:FIG1}(c).
Other fit parameters ($\alpha$, $\lambda_0$, and $\phi$) are shown in Fig. S1 of the Supplemental Material (SM) \cite{SM}.
While $\lambda_0$ remains comparatively small ($<$ 0.02 $\mu$s$^{-1}$), $\lambda$ shows a clear temperature dependence.
It reaches a maximum value at $T$ = 0.55 K and decreases monotonically to a constant value at lower temperatures.
Similarly, $B_\mathrm{int}$ grows and reaches $B_\mathrm{int}$ $\sim$ 56 Oe below $T$ = 0.3 K.

$B_\mathrm{int}$($T$), considered as a measure of an order parameter, indicates the onset of magnetic order below $T$ = 0.55 K.
This is the first direct identification of magnetic order in the $T_\mathrm{o}$ state.
Furthermore, this reveals that superconductivity occurs in the magnetically ordered state, where time-reversal symmetry is broken.
This is a crucial information for determining the SC order parameter.
The previous understanding of the SC phase diagram ascribed the low-field SC state to a (pseudo)spin-singlet even-parity state because time reversal symmetry and fourfold spatial symmetry were assumed to be conserved.
Our findings suggest that this description may be insufficient to fully capture the SC order parameter and open the possibility for a more exotic order \cite{Szabo2023}.
Unfortunately, the pre-established magnetic order above $T_\mathrm{c}$ makes it challenging to separately resolve time-reversal symmetry of the SC order parameter in ZF-$\mu$SR measurements.
A slight suppression of $B_\mathrm{int}$ below $T_\mathrm{c}$ [see the inset of Fig. \ref{fig:FIG1}(c)] may be related to the onset of the SC state.
While this could simply be a change in the hyperfine field due to a depletion of the density of state in the SC state, it could also reflect a subtle coupling between the magnetic and SC order parameters.

\begin{figure}[h!]
\centering
\includegraphics[width=80mm]{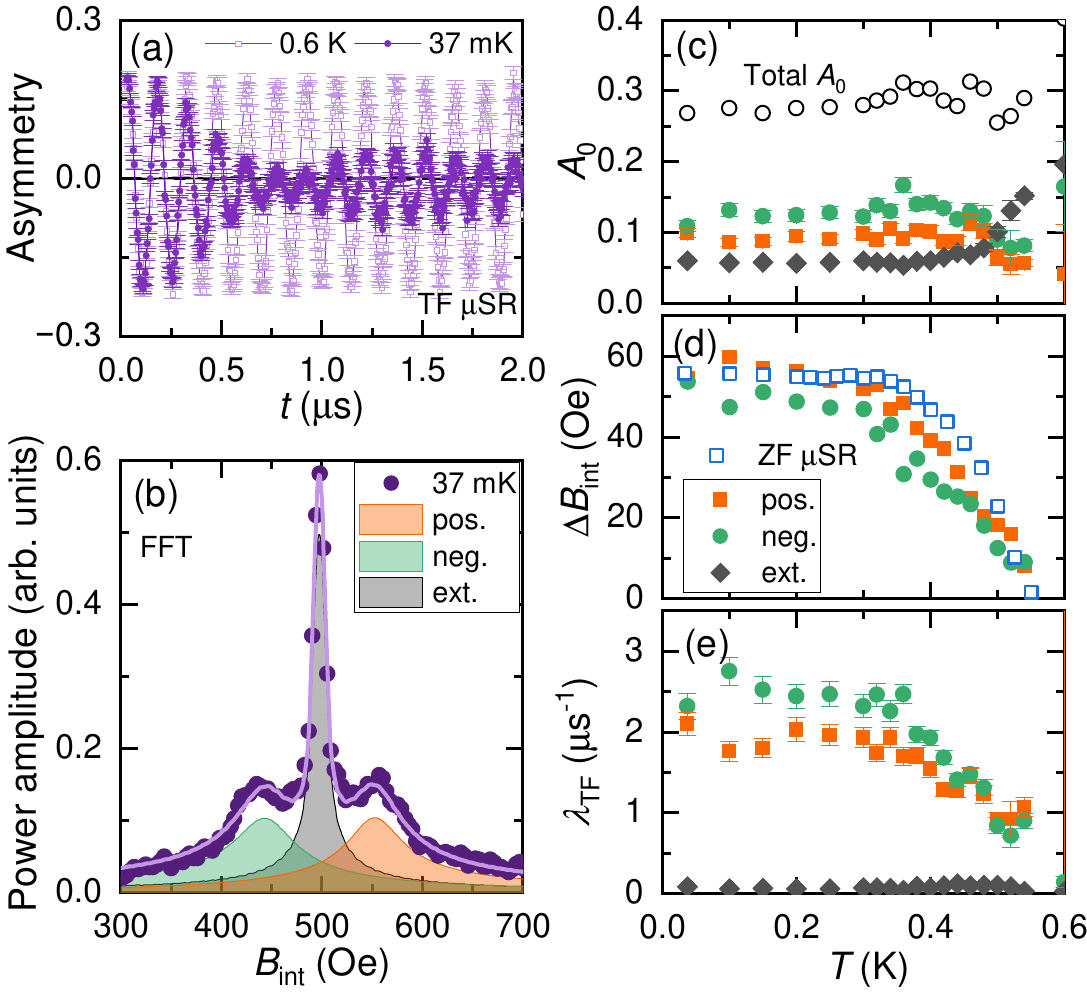}
\caption{
(a) wTF-$\mu$SR time spectra at $T$ = 0.6 K and 37 mK, respectively. (b) FFT power amplitude of the time spectra at $T$ = 37 mK. The solid line denotes the fit given by Eq. \ref{eq:cosine}. Each contribution from the external field (gray) and the positive (orange) and negative (green) internal field are separately shown. Temperature dependence of (c) muon asymmetry, (d) internal field $\Delta B_\mathrm{int}$ = $\lvert B_\mathrm{int}-B_\mathrm{ext}\rvert$, and (e) relaxation rate $\lambda_\mathrm{TF}$. The total sum of the asymmetries (open circle) and the internal field in the ZF $\mu$SR (open blue square) are shown together in (c) and (d), respectively.
}
\label{fig:FIG2}
\end{figure}

We also performed wTF-$\mu$SR experiments to investigate the distribution of the internal local field in more detail.
Figure~\ref{fig:FIG2}(a) presents the TF-$\mu$SR time spectra obtained under an applied field of $B_\mathrm{ext}$ = 500~Oe above and below $T_\mathrm{o}$ and $T_\mathrm{c}$, respectively.
The data at $T$ = 37~mK were taken in the field-cooling (FC) condition.
We see a clear suppression of the oscillation amplitude and a beating behavior at $T$ = 37~mK.
The corresponding FFT power spectrum in Fig.~\ref{fig:FIG2}(b) identifies two broad peaks closely spaced on either side of the sharp central peak.
For the given TF-$\mu$SR experimental configuration, two additional satellite contributions denote the internal field distributions projected along the $c$-axis.

The wTF-$\mu$SR time spectra obtained at various temperatures were fitted to the empirical function written as
\begin{align}
P(t) &= \sum_{i=\mathrm{pos.,neg.}} A_{\mathrm{int},i}\exp\left[ -\lambda_{\mathrm{TF},i} t \right] \cos(\gamma_{\mu}B_{\mathrm{int},i}t + \varphi_{\mathrm{int},i}) \nonumber \\
&+ A_\mathrm{ext} \exp\left[ -\lambda_\mathrm{TF,ext} t \right] \cos(\gamma_{\mu}B_\mathrm{ext}t + \varphi_\mathrm{ext}),
\label{eq:cosine}
\end{align}

where $A$ is the initial muon asymmetry of each component.
As illustrated in Fig. \ref{fig:FIG2}(b), the model describes the entire FFT spectrum quite well.
The obtained fit parameters are summarized in Fig. \ref{fig:FIG2}(c-e).  
The increases in the muon asymmetries, the internal field shifts  $\Delta B_\mathrm{int}$ = $\lvert B_\mathrm{int}-B_\mathrm{ext}\rvert$, and the broadenings of two contributions appear simultaneously below $T_ \mathrm{o}$, confirming the presence of magnetic order.
The fraction of the muon asymmetries for the emerging internal fields below $T_o$ is more than 70 \% of the total $A_0$ [Fig. \ref{fig:FIG2}(c)], reflecting the bulk nature of the magnetic order.  
Furthermore, Fig. \ref{fig:FIG2}(d) compares $\Delta B_{int}$ determined in wTF-$\mu$SR with $B_\mathrm{int}$ in ZF-$\mu$SR.
Their temperature-dependent behavior and absolute values are very similar.
This implies that the same magnetic structure is responsible for the opposite pair of the internal field shifts seen in wTF-$\mu$SR.
In particular, the similar size of $\Delta B_{int}$ in ZF- and wTF-$\mu$SR indicates a comparable in-plane and $c$-axis component of the internal field, which may suggest a weakly anisotropic magnetic structure.

\begin{figure}[h]
\centering
\includegraphics[width=80mm]{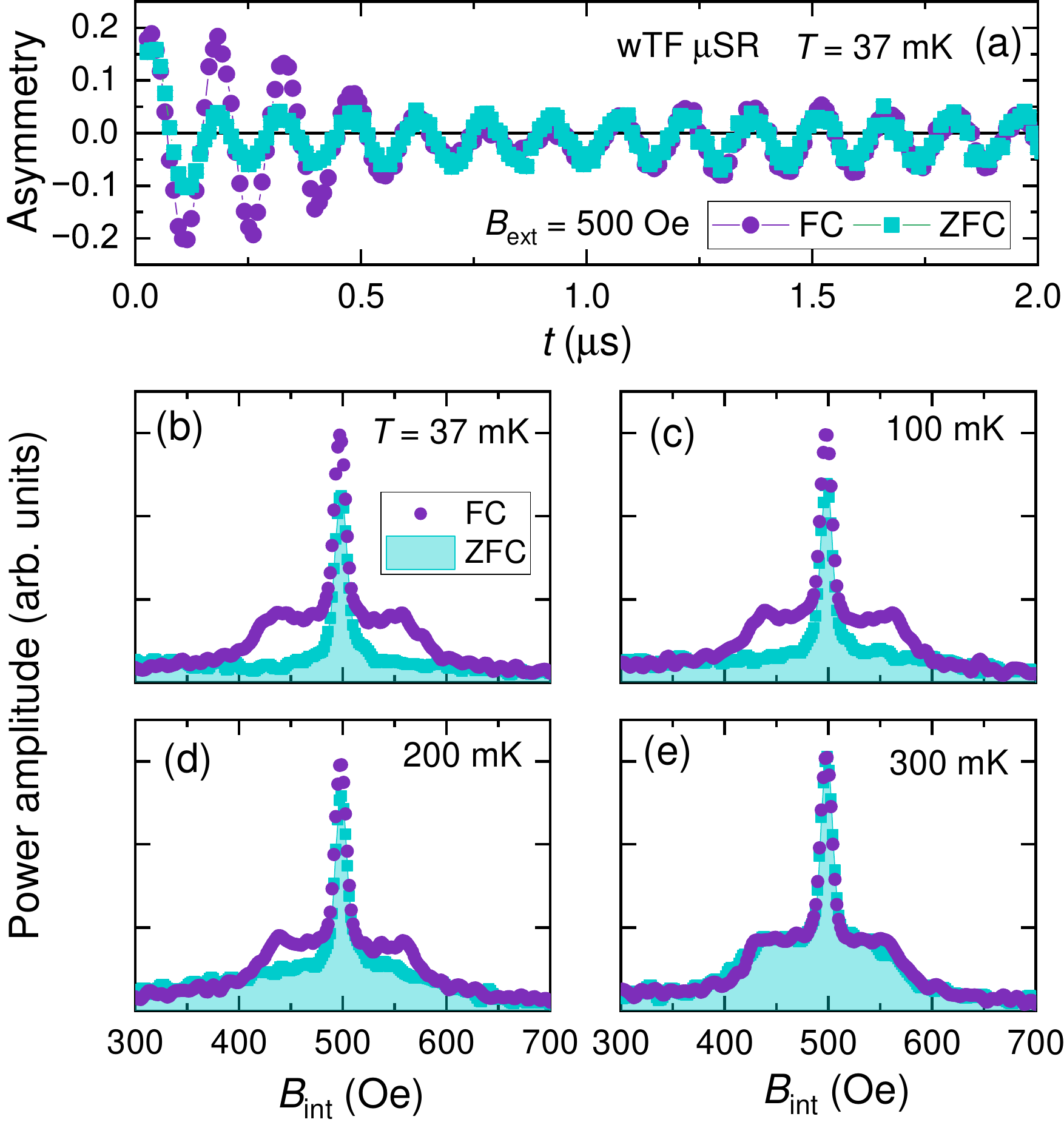}
\caption{wTF-$\mu$SR time spectra at $T$ = 37 mK under the FC and ZFC conditions, respectively. FFT power amplitude of the FC and ZFC data at (b) $T$ = 37 mK, (c) 100 mK, (d) 200 mK, and (e) 300 mK.}
\label{fig:FIG3}
\end{figure}

The above-mentioned wTF-$\mu$SR measurements, performed in the field-cooling (FC) condition, do not show a clear signature of the SC transition at $T_\mathrm{c}$ = 0.3 K.
This is rather unexpected, since heat-capacity measurements confirmed a bulk SC transition at $T_\mathrm{c}$ (see SM \cite{SM}), so the formation of the SC vortex lattice and the resulting internal field distribution would lead to an additional increase of the muon relaxation rate.
On the other hand, the SC relaxation rate, which is related to the SC penetration depth, tends to be proportional to $T_\mathrm{c}$ and inversely related to the effective mass \cite{uemura1991,Sonier2000,MacLaughlin2002}.
The low $T_\mathrm{c}$ and the strongly enhanced effective mass in CeRh$_2$As$_2$ could result in a small SC relaxation rate, which could not be easily resolved in the presence of local magnetism \cite{Uemura1989}.
If so, TF-$\mu$SR measurements in the zero-field cooling (ZFC) condition are highly recommended to test the existence of bulk superconductivity, since a disordered vortex lattice, which is a typical of the ZFC protocol, will increase the SC relaxation rate.

Figure~\ref{fig:FIG3}(a) compares TF-$\mu$SR time spectra obtained from FC (shown in Fig.~\ref{fig:FIG2}(a)) with those obtained from ZFC.
For the ZFC protocol, we cooled down the samples to the lowest temperature at zero field and applied a field of $B_\mathrm{ext}$ = 500 Oe. 
We see a stronger suppression of the muon asymmetry oscillation at $t$ $<$ 0.5 $\mu$s$^{-1}$ after ZFC.
This difference at various temperatures is highlighted in the series of FFT spectra shown in Fig. \ref{fig:FIG3}(b-e).
The characteristic local field distribution in FC remains almost unchanged below $T$ = 0.3 K as shown earlier.
In contrast, the ZFC data vary strongly with temperature:
At the lowest temperature ($T$ = 37 mK) [Fig. \ref{fig:FIG3}(b)], the central weight is significantly reduced and the satellite feature is absent, meaning that a large fraction of the spectral weight is broadly distributed and therefore barely seen.
As the temperature is increased [Fig. \ref{fig:FIG3}(c-d)], the invisible broad feature grows into a hump, then finally turns into the same feature of the FC data above $T_\mathrm{c}$ [Fig. \ref{fig:FIG3} (e)].
The different muon relaxation behavior in the ZFC and FC conditions below $T_\mathrm{c}$  directly proves the existence of bulk superconductivity.
Hence, our $\mu$SR experiments successfully identify two anomalies, a magnetic transition at $T_\mathrm{o}$ and a SC transition at $T_\mathrm{c}$.
This fully agrees with thermodynamic measurements of heat capacity (see SM \cite{SM}) and thermal expansion \cite{khanenko}.

\begin{figure}[h]
\centering
\includegraphics[width=80mm]{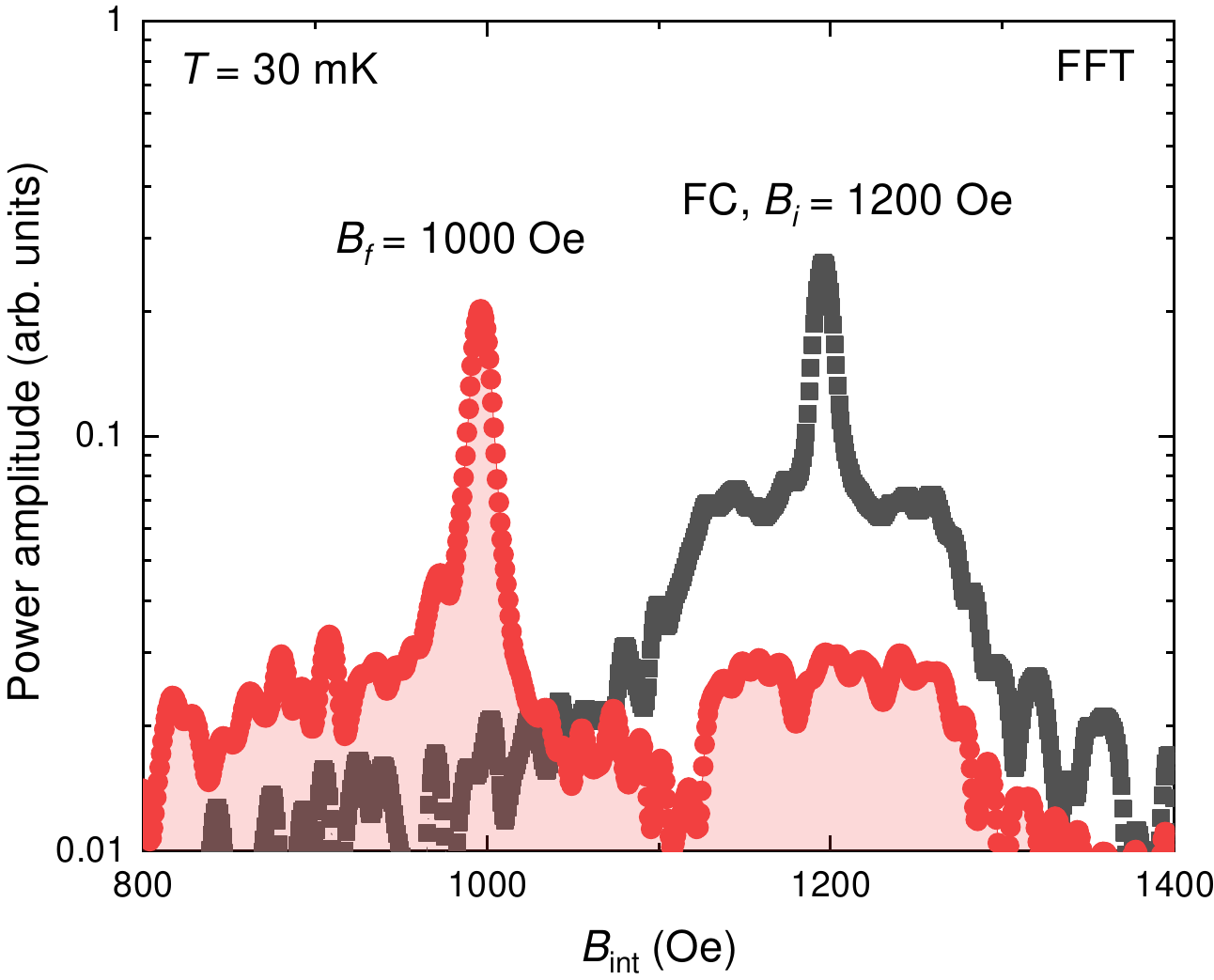}
\caption{(a) Fast Fourier Transform spectra of the TF $\mu$SR data in $T$ = 30 mK. The measurements were carried out for $B_\mathrm{i}$ = 1200 Oe in the FC condition (black square). Then the measurements were repeated after the external fields was subsequently reduced to $B_\mathrm{f}$ = 1000 Oe (red circles).}
\label{fig:FIG4}
\end{figure}

Since the magnetic order persists in the bulk SC state, we should elaborate the nature of the coexistence of superconductivity and magnetism.
This motivated us to perform some vortex-pinning experiments.
These experiments are designed to probe local magnetism in the presence of pinned SC vortices, which occur in a different field from the applied external field under which the data are being taken \cite{Howald2015}.
First, the samples are cooled down to the base temperature ($T$ = 30 mK) across $T_\mathrm{c}$ in an initial applied field, $B_i$ = 1200 Oe.
Subsequently, the initial field is reduced to the final field, $B_f$ = 1000 Oe, while keeping the temperature constant.
This allows us to detect a local field distribution in a bulk-SC region where the pinned vortices are locally present.
Figure \ref{fig:FIG4} shows two FFT spectra of TF-$\mu$SR obtained before and after changing the applied field.
The spectrum under $B_i$ demonstrates the characteristic local field distribution centered on the applied field, $B_i$ (cf. also Fig.~\ref{fig:FIG3}).
When the field is changed to $B_f$, this field distribution, although the amplitude of its spectral weight is reduced, remains at the same position around $B_i$.
This hints at local magnetism in the SC region and provides compelling evidence for the coexistence of magnetism and superconductivity on a microscopic level.

The simultaneous occurrence of magnetism and superconductivity in Ce-based intermetallic compounds is a fundamentally interesting phenomenon, since a single 4$f$ electron for the Ce$^{3+}$ valence state should be responsible for both phases, which are mutually exclusive.
The way in which these phases coexist varies between different materials.
In the archetypal heavy-fermion system of CeCu$_2$Si$_2$, the magnetic and SC phases compete with each other, and, hence, are spatially segregated \cite{Feyerherm1997,Stockert2006}.
On the other hand, CePt$_3$Si shows a microscopic coexistence \cite{Amato2005}.
Among these examples, we emphasize that CeRh$_2$As$_2$ is a unique system to realize the microscopic coexistence of magnetism and superconductivity very close to a quantum critical point.
The dual nature of the 4$f$ electrons in CeRh$_2$As$_2$ would potentially involve multiband and a spin-split character\sout{s} of the electronic structures given by the nonsymmorphic symmetry \cite{Hafner2022,Cavanagh2022,wu2023,chen2023}, which gives rise to a situation where the contribution to superconductivity and magnetism varies in different parts of the Fermi surface.

While previous NQR studies found a clear signature of antiferromagnetic (AFM) order below $T_\mathrm{N}$ $\sim$ 0.3 K \cite{Kibune2021}, thermodynamic characterizations, such as heat capacity and thermal expansion measurements, identified only two anomalies at $T_\mathrm{c}$ and $T_ \mathrm{o}$, questioning the existence of the $T_\mathrm{N}$ order \cite{Hafner2022}.
One possible explanation would be that $T_\mathrm{N}$ happens to coincide with $T_\mathrm{c}$.
However, our $\mu$SR results clearly rule out this scenario, as we could not find any additional magnetic anomaly at $T_\mathrm{N}$.
In addition, it is puzzling to understand the absence of a magnetic signal at $T_ \mathrm{o}$ in d.c. magnetization and a.c. magnetic susceptibility measurements at low frequencies \cite{khim2021} in contrast to our finding in $\mu$SR.
The inconsistent results from the different magnetic probes may lead to the possibility of a dynamic nature of magnetic order.
Comparing the gyromagnetic ratios of the probes ($\gamma_{\mu}$ = 135.5 MHz/T vs. $\gamma_{^{75}\mathrm{As}}$ = 7.292 MHz/T), $\mu$SR has a faster experimental timescale than $^{75}$As-NQR.
In this scenario, $\mu$SR can see a coherent internal field given by quasi-static yet fluctuating moments in a timescale of $\tau \sim 1/\gamma_{\mu}$  below $T_ \mathrm{o}$, while the d.c.-like magnetization and $^{75}$As-NQR can hardly see it on a longer timescale, since $\tau \ll 1/\gamma_{^{75}\mathrm{As}}$.
If slowing down of the magnetic fluctuations occurs at a further lower temperature, NQR would probe the internal field below $T_\mathrm{N}$ as a freezing temperature.
We therefore suggest that a dynamic nature of magnetic order and its unique evolution with temperature can reasonably explain the distinct manifestation of the $T_\mathrm{o}$ transition with respect to the probing timescales.

We now turn to the mutual influence of the SC and $T_ \mathrm{o}$ phases.
Taking into account the respective transition temperatures, the relevant energy scales for each phase are separated and, thus, the coupling between them should be weak \cite{Semeniuk2023}.
Nevertheless, the evolution of the magnetic phase seems to be correlated with the onset of superconductivity:
the nearly temperature-independent behavior of $\lambda$($T$) and $B_\mathrm{int}$($T$) [Fig.~\ref{fig:FIG1}(d)], and the slowing down of magnetic fluctuations appear simultaneously below $T_\mathrm{c}$.
This indicates that the magnetic response changes below $T_\mathrm{o}$ until the onset of superconductivity.
Or, conversely, the SC state prevents the further development of magnetic order.
Under external magnetic field, the NMR studies detected the AFM signature only in the low-field SC state, suggesting a close relationship between the magnetism and superconductivity \cite{Ogata2023}.
The implications of this coincidence need to be investigated by means of further in-depth magnetic characterizations.

Our findings of the magnetic order in the $T_\mathrm{o}$ state does not necessarily exclude the proposed itinerant order of multipolar Ce-4$f$ moments, namely the quadrupolar density wave (QDW) order \cite{Hafner2022} or coexisting dipolar and multipolar order \cite{Schmidt2024}.
While the dipole moments are conventionally given by the doublet ground state for the local CEF, the quasi-quartet CEF configuration still allows multipolar degrees of freedom.
Indeed, the magnetic phase diagrams of $T_\mathrm{o}$($B$) show a significant deviation from the expected behavior in an ordinary AFM system and their similarity to typical Ce-based quadrupolar systems has been discussed \cite{Effantin1985,Goto2009,Hidaka2022}.
We also point out that the unusual field-induced behaviors may be related to local inversion symmetry breaking and the associated antisymmetric Dzyaloshinskii-Moriya (DM) interaction, which was evidenced in the noncentrosymmetric CePt$_3$Si \cite{Kaneko2012}.
Finally, we believe that elucidating the origin of the $T_\mathrm{o}$ phase is closely related to understanding the non-Fermi liquid behavior in the normal state \cite{Hafner2022}.
The observed strongly field-robust non-Fermi liquid behavior may open the possibility that quantum criticality is associated with both the electronic multipolar degrees of freedom as well as the spin-orbit related interactions given by the nonsymmorphic crystal structure \cite{khanenko}.

\textit{Conclusions} - In this work, we presented zero-field and transverse-field $\mu$SR studies of the magnetic and SC properties in CeRh$_2$As$_2$ single crystals at low temperatures.
For the first time, we identify a spontaneous internal field below $T_\mathrm{o}$ = 0.55 K, revealing the magnetic origin of the $T_\mathrm{o}$ order.
The coincidence of the SC transition and a fully established magnetic order may imply a close relationship between the SC state and the $T_ \mathrm{o}$ order.
Furthermore, we find evidence for a microscopic coexistence of itinerant magnetism with bulk superconductivity, which marks a rare example of such coexistence occurring very close to a quantum critical point.
We also discuss the possible dynamic nature of the magnetic order to explain the discrepancy between magnetic measurements on a different timescales.
Our results open the possibility that the $T_ \mathrm{o}$ phase involves both dipole and higher-order Ce-4$f$ moment degrees of freedom and account for the unusual non-Fermi liquid behavior.

\begin{acknowledgments}
This work was supported by the Max Planck Society and the Deutsche Forschungsgemeinschaft (DFG, German Research Foundation) - KH 387/1-1.
The authors thank K. Ishida, S. Kitagawa, M. Baenitz, E. Hassinger, G. Zwicknagl, P. Coleman, P. Thalmeier, and R. Sarkar for useful discussions.

\end{acknowledgments}

%


\clearpage
\widetext
\begin{center}
\textbf{\large Supplemental Materials: Coexistence of local magnetism and superconductivity in the heavy-fermion CeRh$_2$As$_2$ revealed by $\mu$SR studies}
\end{center}
\setcounter{equation}{0}
\setcounter{figure}{0}
\setcounter{table}{0}
\setcounter{page}{1}
\makeatletter
\renewcommand{\theequation}{S\arabic{equation}}
\renewcommand{\thefigure}{S\arabic{figure}}
\renewcommand{\bibnumfmt}[1]{[S#1]}
\renewcommand{\citenumfont}[1]{S#1}

\section{Temperature dependence of the fit parameters of $\alpha$, $\lambda_0$, and $\phi_0$ used in \NoCaseChange{Eq. (1)}}

\begin{figure}[h]
\centering
\includegraphics[width=85mm]{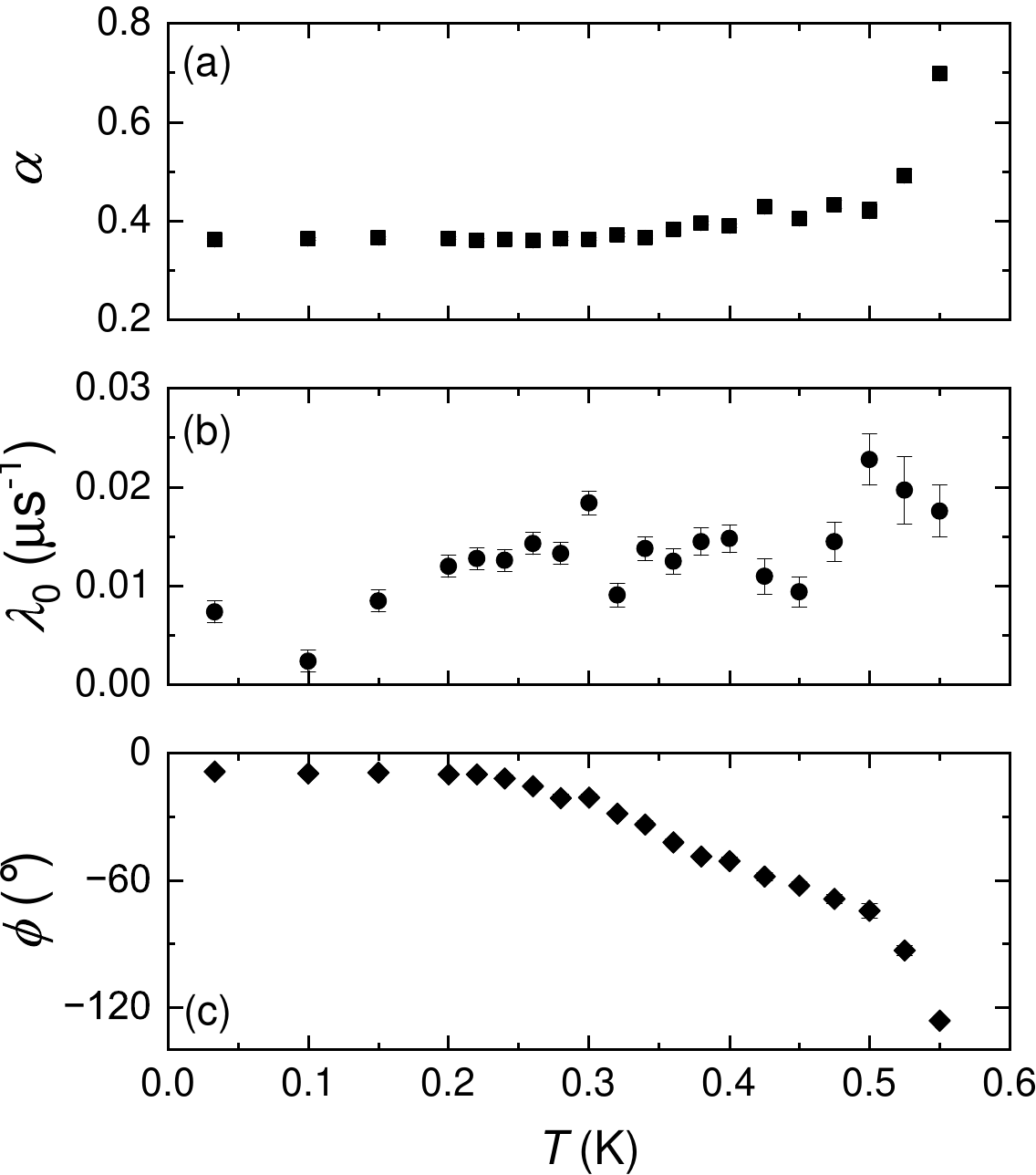}
\caption{Temperature dependence of (a) $\alpha$, (b) $\lambda_\mathrm{0}$, and (c) $\phi$ used to fit the ZF-$\mu$SR time spectra in the main text (Eq. 1).}
\label{fig:FIGS1}
\end{figure}

\pagebreak

\section{Specific-heat measurements}
\begin{figure}[h]
\centering
\includegraphics[width=120mm]{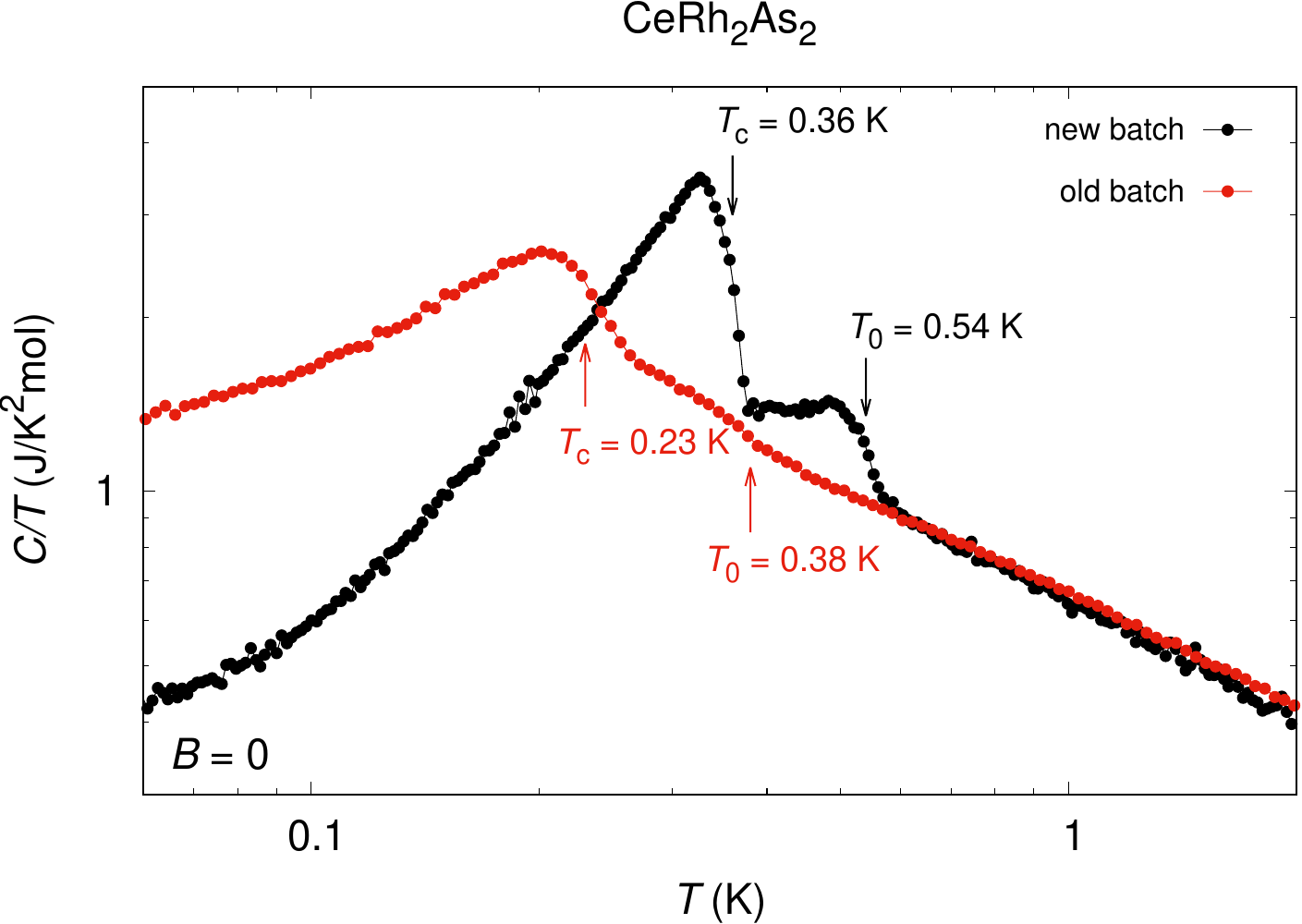}
\caption{Temperature dependent specific heat measured in the ``old'' (red symbol) and ``new'' (black symbol) batches, respectively.}
\label{fig:FIGS2}
\end{figure}

Specific-heat measurements were performed in order to evaluate the quality of the samples used in the $\mu$SR experiments and to determine the transition temperatures for the SC and $T_\mathrm{o}$ phases.
Figure \ref{fig:FIGS2} shows the temperature dependence of the specific heat divided by the temperature ($C$/$T$) of two samples of different quality.
The sample from an ``old batch'' shows a jump at $T_\mathrm{c}$ = 0.2~K and a broad hump feature below $T_\mathrm{o}$ = 0.4~K.
In the ``new batch'' sample discussed in the main text, both transitions are much sharper and $T_\mathrm{c}$ = 0.3 K and $T_\mathrm{o}$ = 0.55~K are determined, indicating a higher quality with less imperfections.
Detailed characterizations on this sample was reported in our previous work \cite{Semeniuk2023}.
We note that the transition temperatures determined from specific-heat data are in good agreement with the results of the $\mu$SR experiments.
For completeness, below we present also the $\mu$SR measurements on the "old batch" samples are presented in the following section.

\pagebreak

\section{$\mu$SR measurement results on ``old-batch'' \NoCaseChange{CeRh$_2$As$_2$} crystals}

\begin{figure}[h!]
\centering
\includegraphics[width=150mm]{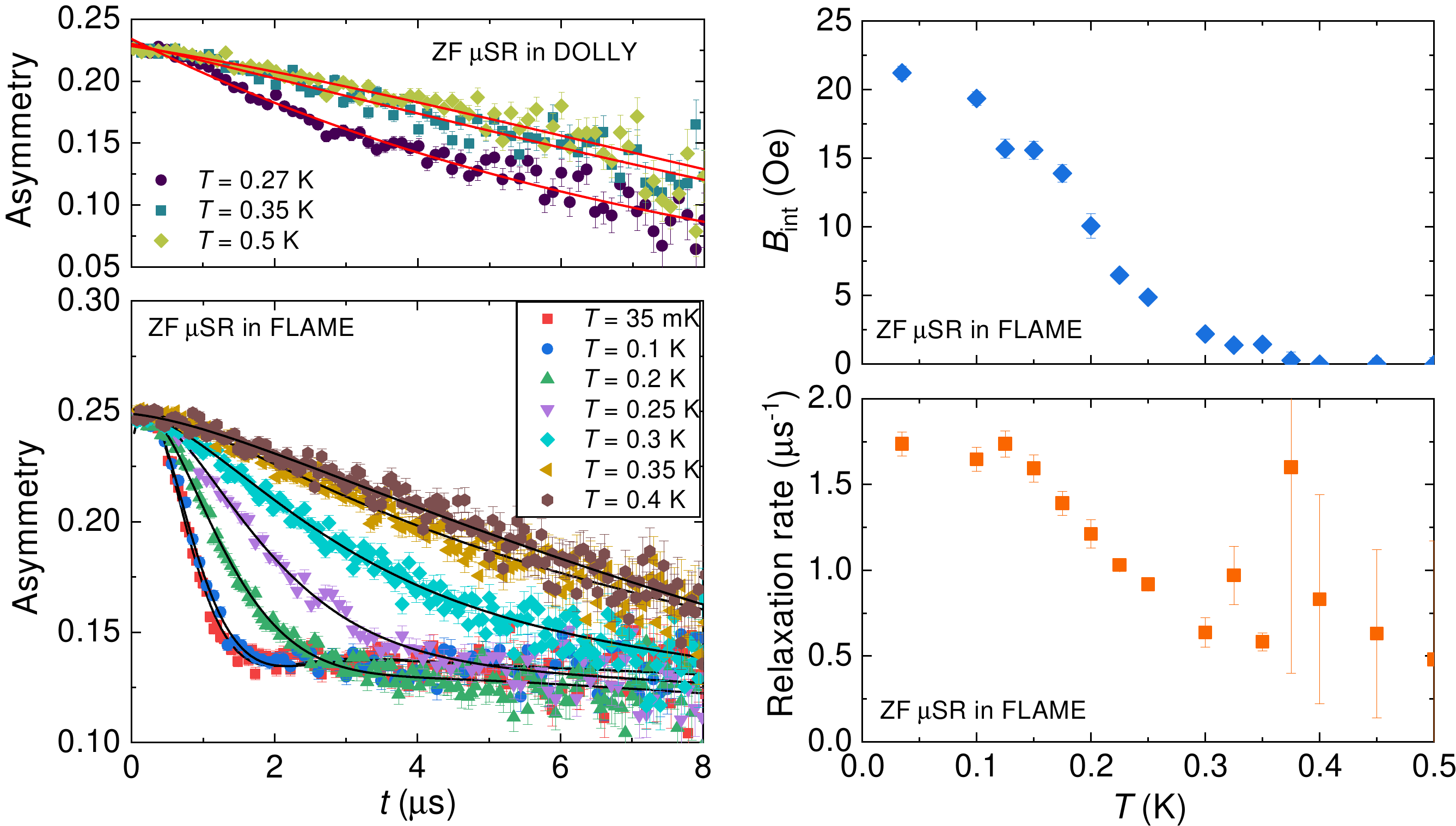}
\caption{(a) ZF-$\mu$SR time spectra obtained on the DOLLY instrument at $T$ = 0.27, 0.35, and 0.5~K, and (b) on the FLAME instrument down to $T$ = 35~mK. The FLAME data were fitted to the model of Eq. \ref{eq:cosine_ZF}. Temperature dependences of (c) internal field and (d) relaxation rate determined from the model fit.}
\label{fig:FIGS3}
\end{figure}

\subsection{Zero-field $\mu$SR measurements}
The zero-field measurements were carried out on the Dolly and FLAME instruments at the Paul Scherrer Institute (PSI, Villigen, Switzerland).
The measurements on DOLLY down to $T$ = 0.27~K showed that the muon relaxation is enhanced below $T$ $\sim$ 0.3~K.
Subsequent measurements at lower temperatures on FLAME clearly identified the significant loss of the muon asymmetry due to the onset of a magnetic order.
The FLAME data are fitted to a formula written as

\begin{equation}
A(t) = A_{0} e^{-\lambda_{0} t} + A \cos \left( 2\pi\nu t + \frac{\pi\phi}{180} \right) e^{-\lambda t}.
\label{eq:cosine_ZF}
\end{equation}

While an oscillating feature is not clearly seen in the time spectra, the fit results indicate a systematic increase of the oscillation frequency below $T$ = 0.4~K [Fig. \ref{fig:FIGS3}(c)], which is attributed to the $T_\mathrm{o}$ order.
The $T_\mathrm{o}$ transition is broad, reflecting an inhomogeneity of the samples already shown in the specific-heat data [Fig. \ref{fig:FIGS2}].
The internal field at the lowest temperature of $B_\mathrm{int}$ $\sim$ 22 Oe is apparently smaller than that in the quality-improved samples ($B_\mathrm{int}$ $\sim$ 56 Oe) discussed in the main text.
This indicates that $B_\mathrm{int}$ in the $T_0$ phase strongly depends on crystalline disorder or/and impurity.
This could reflect the short-range nature of the magnetic order or the sensitivity of the magnetic structure to small perturbations near a quantum critical point.
Here, we observed no change in the initial asymmetry below $T_\mathrm{o}$.

\pagebreak

\subsection{Weak transverse-field $\mu$SR measurements}

\begin{figure}[h!]
\centering
\includegraphics[width=150mm]{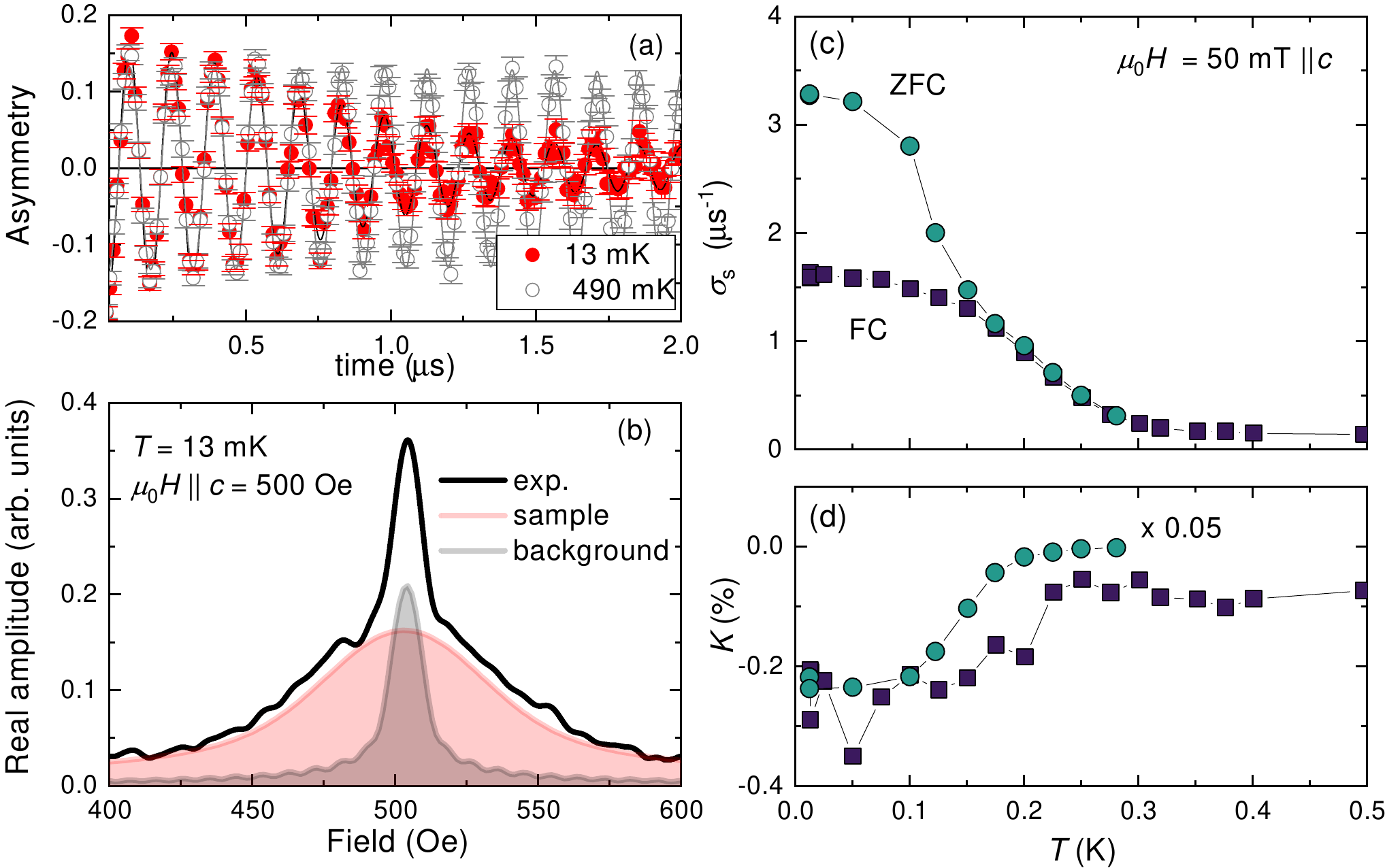}
\caption{(a) wTF-$\mu$SR time spectra under $B_{\mathrm{ext}}$ = 500~Oe at $T$ = 13~mK and 490~mK, respectively. (b) Fast Fourier Transform spectra of the TF $\mu$SR data at $T$ = 13~mK. (c) Temperature dependence of (d) relaxation rate $\sigma_\mathrm{s}$ and muon knight shift $K$ determined in the FC and ZFC conditions, respectively.}
\label{fig:FIGS4}
\end{figure}

Weak transverse-field (wTF) measurements were carried out on the HAL-9500 instrument at the Paul Scherrer Institute (PSI, Villigen, Switzerland).
Figure \ref{fig:FIGS4}(a) shows wTF-$\mu$SR time spectra $A$($t$).
At $T$ = 0.49~K, the signal oscillates with a small relaxation, mainly given by a randomly distributed local field.
At $T$ = 13~mK, in contrast, the oscillation amplitude apparently decreases.
The corresponding FFT power spectra are shown in Fig. \ref{fig:FIGS4}(b).
Here, a single broad contribution is identified to account for local magnetism.
This is distinguished from the two clearly resolved peaks in the ``new batch'' samples shown in the main text.
We analyze $A$($t$) by fitting the two-component cosine function with the Gaussian relaxation which reasonably describes the muon relaxation by static homogeneous internal fields, expressed as

\begin{align}
A(t) &= \sum_{i = s, bg} A_{i}\exp\left[ -\frac{\sigma_{i}^2t^2}{2} \right] \cos(\gamma_{\mu}B_{\mathrm{int},i}t + \varphi_{i}),
\label{eq:eq1}
\end{align}
where $A_i$, $\sigma_{i}$, $B_{int,i}$, and $\varphi_{i}$ is the initial asymmetry, relaxation rate, internal field, and initial phase of the muon-spin ensemble for the sample ($\textit{i}$ = $\textit{s}$) and background contribution ($i = bg$), respectively.

Figure \ref{fig:FIGS4}(c) shows the temperature($T$)-dependent muon relaxation rate $\sigma_{s}(T)$ obtained in the field-cooled (FC) and zero-field-cooled (ZFC) conditions, respectively.
Both $\sigma_{s}$ increase with decreasing $T$ below $\sim$ 0.4~K due to the onset of superconductivity.
On further cooling, they diverge from each other below 0.22~K, with the ZFC data marking a larger value.
At the same temperature, the muon Knight shift, $K$ = ($B_{\mathrm{int},s}$-$B_{\mathrm{int},bg}$)/$B_{\mathrm{int},bg}$, decreases significantly, as shown in Fig. \ref{fig:FIGS4}(d).
These observations point to the bulk nature of superconductivity below $T_\mathrm{c}$ = 0.22~K.

\end{document}